\documentclass[acmlarge,screen]{acmart}

\usepackage[figuresright]{rotating}
\usepackage{tabularx}
\usepackage{enumitem}
\usepackage{multirow}
\usepackage{multicol}
\usepackage{supertabular}

\usepackage{lscape}

\usepackage{tcolorbox}
\usepackage{color}

\usepackage[misc]{ifsym}  % corresponding author

\AtBeginDocument{%
  \providecommand\BibTeX{{%
    \normalfont B\kern-0.5em{\scshape i\kern-0.25em b}\kern-0.8em\TeX}}}

\acmPrice{15.00}
\acmISBN{978-1-4503-XXXX-X/18/06}

\begin{document}
	
% \linespread{2} %修改行距 2倍

\settopmatter{printacmref=false} 
\renewcommand\footnotetextcopyrightpermission[1]{}
	
\graphicspath{{figs/}}
% \graphicspath{{AIPPT/}}

% \title{A Systemic Review of Contactless Physiological Signal Monitoring in Clinical: Applications, Challenges and Future Work}

% \title{A Public Health Large Medical Model via Contactless Physiological Monitoring and Cardiopulmonary Diagnosis}

% \title{A Framework of Large Medical Model for Public Health via Visual Contactless Physiological Monitoring}

\title{A Framework of Large Medical Model for Public Health via Camera-based Physiological Monitoring}

\title{A Framework of Large Medical Model for Public Health via Visual-based Physiological Monitoring}

\title{Large Medical Artificial Intelligence Models for Public Health}

\title{A Large Medical Model based on Visual Physiological Monitoring for Public Health}
\title{AI for Healthcare and Public Health Based on Visual Physiological Monitoring}

\title{A Framework Integrated AI for Public Health with Visual Physiological Monitoring}

\title{A Framework of AI in Public Health with Visual Physiological Monitoring}

\title{A Camera-Based Public Health Framework with Artificial Intelligence}

\title{A Camera-Based Physiological Monitoring and Healthcare Framework Integrating Artificial Intelligence for Public Health}

\title{AI-Enabled Visual Physiological Monitoring: A Sustainable Framework for Health Care}

\title{Camera-based physiological monitoring for public health}

\title{A Camera-Based Physiological Monitoring Framework Integrating Artificial Intelligence for Public Health}

\title{Camera-Based Physiological Monitoring for  Public Health}

\title{An AI-Enabled Framework Within Reach for Enhancing Healthcare Sustainability and Fairness}

% \title{Visual Physiological Monitoring for Healthcare and Public Health}

\author{Bin Huang $^{\textrm{\Letter}}$}
\orcid{0000-0003-0894-5335}
\affiliation{%
	\institution{AI Research Center, Hangzhou Innovation Institute, Beihang University}
	% \streetaddress{99 Juhang Rd., Binjiang Dist.}
	\city{Hangzhou}
	\state{Zhejiang}
	\country{CHN}
}
\email{marshuangbin@buaa.edu.cn}

%\author{Shen Hu}
%\affiliation{%
%	\institution{Department of Obstetrics, The Second Affiliated Hospital of Zhejiang University School of Medicine}
%	% \streetaddress{1511 Jianghong Rd., Binjiang Dist.}
%	\city{Hangzhou}
%	\state{Zhejiang}
%	\country{CHN}}
%\affiliation{%
%	\institution{Department of Epidemiology, The Harvard T.H. Chan School of Public Health}
%	% \streetaddress{677 Huntington Ave}
%	\city{Boston}
%	\state{Massachusetts}
%	\country{USA}}
% \email{2515527@zju.edu.cn}

\author{Changchen Zhao}
\orcid{0000-0002-0546-6016}
\affiliation{%
	\institution{School of Computer Science, Hangzhou Dianzi University}
	% \streetaddress{99 Juhang Rd, Binjiang Dist.}
	% \city{Hangzhou}
	% \state{Zhejiang}
	\country{CHN}
}

\author{Zimeng Liu}
\affiliation{%
	\institution{School of Automation Science and Electrical Engineering, Beihang University}
	% \streetaddress{37 Xueyuan Rd., Haidian Dist.}
	% \city{Beijing}
	\country{CHN}}

\author{Shenda Hong}
\orcid{0000-0001-7521-5127}
\affiliation{%
	\institution{Institute of Medical Technology, Health Science Center of Peking University, and National Institute of Health Data Science, Peking University}
	% \streetaddress{1088 Xueyuan Ave, Nanshan Dist.}
	% \city{Beijing}
	\country{CHN}}

\author{Baochang Zhang}
\affiliation{%
  \institution{Institute of Artificial Intelligence, Beihang University}
  % \city{Beijing}
  \country{CHN}}
 % \email{bczhang@buaa.edu.cn}

\author{Hao Lu}
\affiliation{%
	\institution{Information Hub, Hong Kong University of Science and Technology}
	% \city{Beijing}
	\country{CHN}}
%\email{hlu585@connect.hkust-gz.edu.cn} 

\author{Zhijun Liu}
% \orcid{0000-0003-4150-5470}
\affiliation{%
		\institution{Department of Biomedical Engineering, City University of Hong Kong}
		% \streetaddress{1088 Xueyuan Ave, Nanshan Dist.}
		% \city{Hongkong}
		\country{CHN}}
%\email{zijunliu4-c@my.cityu.edu.hk}

\author{Wenjin Wang $^{\textrm{\Letter}}$}
\orcid{0000-0001-7832-5444}
\affiliation{%
	\institution{Department of Biomedical Engineering, Southern University of Science and Technology}
	% \streetaddress{1088 Xueyuan Ave, Nanshan Dist.}
	% \city{Shenzhen}
	\country{CHN}}
	\email{wangwj3@sustech.edu.cn}

%\author{Yuan-ting Zhang $^{\textrm{\Letter}}$}
%\orcid{0000-0003-4150-5470}
%\affiliation{%
%	\institution{Department of Biomedical Engineering, City University of Hong Kong}
%	% \streetaddress{1088 Xueyuan Ave, Nanshan Dist.}
%	% \city{Hongkong}
%	\country{CHN}}
%\email{ytzhang@cuhk.edu.hk}

\author{Hui Liu $^{\textrm{\Letter}}$}
\orcid{0000-0002-6233-3973}
\affiliation{%
	\institution{Institute of Medical Information, Chinese Academy of Medical Sciences \& Peking Union Medical College}
	% \streetaddress{1088 Xueyuan Ave, Nanshan Dist.}
	% \city{Beijing}
	\country{CHN}}
    \email{liuhui@pumc.edu.cn}
\renewcommand{\shortauthors}{Trovato and Tobin, et al.}


\begin{abstract}
	\begin{center}
		\begin{tcolorbox}[colback=gray!10, %gray background
			colframe=black, % black frame colour
			width=16cm, % Use 8cm total width,
			arc=1mm, auto outer arc,
			boxrule=0.5pt,]
			
			\textbf{Abstract:} 

			Good health and well-being is among key issues in the United Nations 2030 Sustainable Development Goals. The rising prevalence of large-scale infectious diseases and the accelerated aging of the global population are driving the transformation of healthcare technologies. In this context, establishing large-scale public health datasets, developing medical models, and creating decision-making systems with a human-centric approach are of strategic significance. Recently, by leveraging the extraordinary number of accessible cameras, groundbreaking advancements have emerged in AI methods for physiological signal monitoring and disease diagnosis using camera sensors. These approaches, requiring no specialized medical equipment, offer convenient manners of collecting large-scale medical data in response to public health events. Therefore, we outline a prospective framework and heuristic vision for a camera-based public health (CBPH) framework utilizing visual physiological monitoring technology. The CBPH can be considered as a convenient and universal framework for public health, advancing the United Nations Sustainable Development Goals, particularly in promoting the universality, sustainability, and equity of healthcare in low- and middle-income countries or regions. Furthermore, CBPH provides a comprehensive solution for building a large-scale and human-centric medical database, and a multi-task large medical model for public health and medical scientific discoveries. It has a significant potential to revolutionize personal monitoring technologies, digital medicine, telemedicine, and primary health care in public health. Therefore, it can be deemed that the outcomes of this paper will contribute to the establishment of a sustainable and fair framework for public health, which serves as a crucial bridge for advancing scientific discoveries in the realm of AI for medicine (AI4Medicine).
			
			% Finally, an in-depth analysis is conducted to provide insight into the notable role that PHLMM plays in two distinct typical public health scenarios, namely the response to the COVID-19 pandemic and the management of chronic diseases of the elderly.
			
			% The outcomes of this perspective provide a framework large medical model for public health, which construct a  bridge of scientific discoveries in AI for medicine.
			
			% Firstly, due to the fact that VCPM requires only an off-the-shelf camera and enables contactless measurement, it boasts cost-effectiveness and user-friendly passive physiological monitoring. This provides crucial technological support for constructing large-scale public health datasets. 
			
			% To comprehensively capture the variations in these parameters in real-time, we introduce in this prospective paper a public health medical model system utilizing Visual Contactless Physiological Monitoring (VCPM) technology. Firstly, VCPM enables the convenient, continuous, and real-time monitoring of multiple vital signs, thereby furnishing more accurate and comprehensive data. This capability facilitates the early detection and assessment of physiological abnormalities in patients.
			
		\end{tcolorbox}
	\end{center}

  % \textbf{Keywords:} Contactless physiological monitoring; Deep Learning; Telemedicine; Telehealth; Video-based monitoring;
\end{abstract}

\maketitle

%% todo: 只可有一级子标题

\section{Introduction}

% 大背景

Good health and well-being are among the United Nations' Sustainable Development Goals (SDGs) for the 2030 Agenda; however, they are challenged by the following two major public health issues. The first is the surge in non-communicable chronic diseases caused by the accelerating global aging population, such as the long-term monitoring of elderly patients with cardiovascular diseases. The another is the outbreak and monitoring of global pandemic infectious diseases, such as Disease X or COVID-19 pandemic. Personal health monitoring technologies \cite{hanson2022lancet} play a crucial role in tackling the challenges posed by population aging and large-scale infectious diseases \cite{frieden2023road}. These technologies enable continuous monitoring of vital signs and health parameters, allowing for early detection and intervention of health issues in the elderly, which is essential for maintaining their quality of life and independence \cite{WHOPHCframework2022, croke2024primary, henderson2023understanding, kasper2023rethinking}. Additionally, during outbreaks of infectious diseases, personal health monitoring devices can provide real-time data on symptoms and disease progression, facilitating timely medical responses and reducing the spread of infections. By leveraging advanced sensors and data analytics, these technologies contribute significantly to public health management and the creation of resilient healthcare systems.

% 具体化视觉传感器在健康监测中的作用。 1）传感器的数量，2）可以那些事情 3）为什么要用视觉传感器
But how to develop personal health monitoring technologies to cover a broader population? The ubiquitous visual or camera sensor will be a key component in achieving this goal, such as webcam, smartphone and tablets built-in camera. In 2024, the global number of smartphone users is up to be 4.88 billion, increasing to 6.38 billion by 2029, with a global smartphone usage rate reaching 75\% \cite{how-many-phones2024}. In addition, it has been demonstrated that cameras can be utilized for physiological information monitoring and disease diagnosis \cite{huang2023challenges, lee2023lstc, du2023dual, zhao2023learning, maurya2021non, liu2023contactless, ALNAGGAR2023120135, fawzy2023clinical, mathew2023remote, frey2022blood, curran2023camera, hwang2024phase, rizas2022smartphone, gruwez2023smartphone, khan2023remote, gruwez2024real, fernstad2024validation}, with both contact and non-contact options available. Therefore, the widespread availability of camera sensors can be utilized to promote the development of non-invasive and accessible personal health monitoring systems. This approach can facilitate early detection of health issues, personalized health management, and efficient allocation of healthcare resources, ultimately enhancing the overall well-being of a broader population. For instance, from 2023 to 2024, the official Journals of the European Heart Rhythm Association publish several clinical researches on atrial fibrillation (AF) screening based on smartphone's cameras photoplethysmography (PPG) \cite{fernstad2024validation, gruwez2024real, hermans2023accuracy, gruwez2023smartphone}. These pioneering studies demonstrates the use of cameras for cardiovascular disease monitoring and visual-based disease monitoring solutions for public health. 
Consequently, the Camera-Based Public Health (CBPH) solution can not only reduce healthcare costs and optimize resource allocation \cite{haug2023artificial, arora2023value}, but also decrease cardiovascular disease mortality rates in low- and middle-income countries (LMICs), thereby promoting sustainable development.

\begin{figure}
	\centering
	\includegraphics[width = 16.0cm]{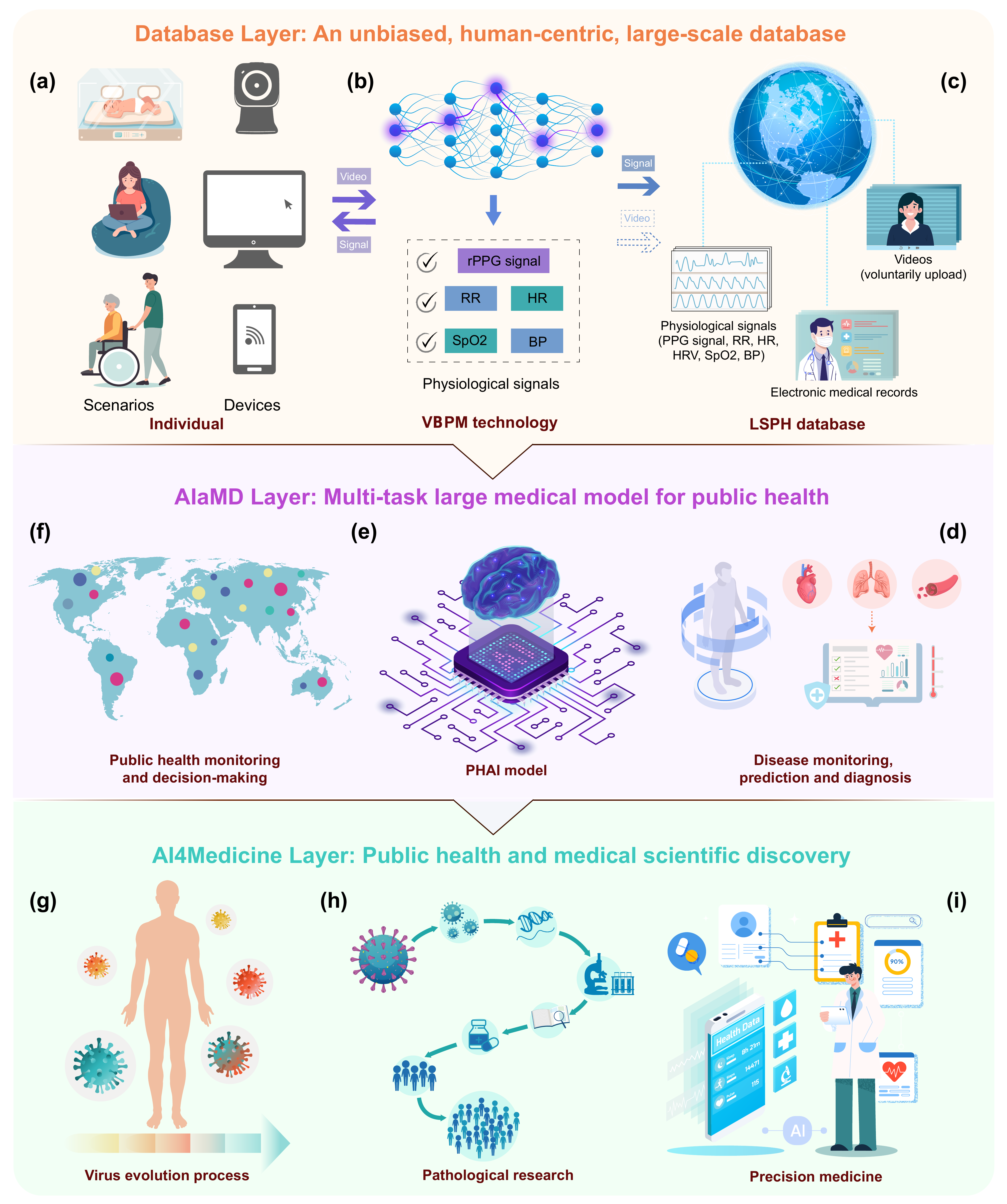}
	\caption{A landscape of the CBPH framework. CBPH consists of three layers: the database, AIaMD and AI4Medicine layers. \textbf{a}, Data collection from the entire population utilizing camera sensors at hand. \textbf{b}, Extraction of physiological information using VBPM technologies. \textbf{c}, The constructed LSPH database, comprising 1) user-uploaded raw video data (voluntarily); 2) physiological information; 3) diagnosis information of patients. \textbf{d}, Monitoring, prediction and diagnosis of some specific diseases, such as AF, hypertension, and vascular aging. \textbf{e}, Development of a multi-task public health AI model. \textbf{f}, Global public health monitoring and decision-making in response to emergency events, such as COVID-19 or Disease X. % based on LSPH database, PHAI model and explanable AI technologies. 
	\textbf{g}, Considering the distinction in various populations (e.g., the elderly and neonates) and those with different underlying diseases, the comprehensive studies on the evolution of diseases provide support for the formulation of personalized treatment plans. \textbf{h}, Medical scientific discoveries and pathological research based on large-scale data and model. \textbf{i}, Discovery of novel digital biomarkers and endpoints for evidence-based medicine and precision medicine.
		% An unbised and human-centric LSPH database is to be constructed in the databasa layer. 
	}
	\label{Fig:PHLMM_overview}
\end{figure}

In the Perspective, we propose a CBPH framework with visual physiological monitoring. As illustrated in Fig. \ref{Fig:PHLMM_overview}, the CBPH system consists of three layers, namely the database, AI as medical device (AIaMD) and AI4Medicine layers. Firstly, an unbiased, human-centric, and large-scale public health (LSPH) database is constructed with visual-based physiological monitoring (VBPM) technologies. Since it relies solely on the use of an off-the-shelf camera, the physiological information across populations and life-cycle can be easily collected from multi-scenarios. Thus, LSPH is a database that covers the public health information of global-time and global-space. Secondly, the multi-task public health AI (PHAI) fundamental model, which is an Large Medical Model (LMM) for healthcare, is designed as AIaMD for disease monitoring, prediction and diagnosis based on LSPH database. Meanwhile, explainable AI (XAI) technologies are employed to enhance the transparency and trustworthiness of the PHAI model, and are used to discover novel digital biomarkers for disease diagnosis. Finally, in the AI4Medicine layer of CBPH, the PHAI fundamental model can be utilized for various downstream tasks, such as medical scientific discoveries and pathological research. Overall, the database and model offer a profound understanding of disease development mechanisms and facilitate public health. Through the analysis of extensive time-series physiological data from the public, CBPH can unveil new disease features and patterns, such as discovering the pattern of physiological changes during SARS-CoV-2 infection, thus enabling the customization of treatment plans and providing precise guidance for pathological diagnosis.

% In the PHLMM system, the public can easily utilize accessible camera devices such as smartphones and webcams to establish personal baseline physiological data through VBPM technology. Subsequently, leveraging the baseline physiological information from daily health activities and physiological data from the onset of a particular disease, PHLMM can be employed to predict the timeline of the onset of certain illnesses, such as cardiovascular diseases (CVDs), and monitor the evolution of diseases, among other applications. Furthermore, as PHLMM can acquire extensive physiological data from voluntarily participating individuals, it can be used to establish LMMs and monitoring \& decision-making systems for medical scientific discoveries and public health emergency, including response to epidemics such as the COVID-19 pandemic \cite{henderson2023understanding}.

Physiological parameters are crucial indicators of an individual's physiological status and overall health. Regular and systematic monitoring of vital signs is a routine in healthcare settings, ranging from hospitals and clinics to home-based nursing. The abnormal fluctuations in these physiological parameters are closely associated with the evolutionary processes of numerous diseases within the human body. Moreover, these physiological parameters are critical in emergency care, preventing complications and guiding clinical decision-making. In the CBPH system, physiological information, including PPG signal \cite{lee2023lstc, du2023dual}, heart rate (HR) \cite{li2023non, zhao2023learning, jorge2022non}, respiration rate (RR) \cite{maurya2021non, liu2023contactless, ALNAGGAR2023120135}, oxygen saturation (SpO$_2$) \cite{fawzy2023clinical, mathew2023remote, hoffman2022smartphone} and blood pressure (BP) \cite{frey2022blood, curran2023camera, hwang2024phase}, can be inferred from skin videos by VBPM technologies \cite{huang2023challenges, mcduff2023camera, diao2022video, wang2021contactless}. Moreover, recent research demonstrates that the physiological information can be employed to predict and diagnose diseases, such as COVID-19 infection \cite{hirten2021use, fawzy2023clinical, clarke2023factors, mitratza2022performance, mol2021heart, hasty2021heart}, AF detection \cite{rizas2022smartphone, ding2023log, guo2021photoplethysmography, gruwez2023smartphone, hermans2023accuracy}, hypertension \cite{frey2022blood, mejia2021classification}, diabetes \cite{lan2023hdformer, avram2020digital}, vascular aging \cite{charlton2022assessing, shin2022photoplethysmogram} and other cardiovascular diseases \cite{lan2023performer, weng2023predicting, alastruey2023arterial, charlton2022wearable, JAVAID2022100379, nie2024review}.

% Then, PHLMM is poised to play a crucial role in advancing Universal Health Coverage (UHC) \cite{world2023assessing, pablos2023universal}. As AI software, it relies exclusively on existing smart devices and network infrastructure, thereby ensuring greater accessibility and opportunities for individuals across various geographical locations and socioeconomic statuses. In addition, through the public participation, PHLMM not only expeditiously acquires large-scale data for responding to public health emergencies but also assists healthcare professionals and policymakers in comprehending novel viruses or diseases through AI models for data analysis. It can also facilitate the swift formulation of response plans. Therefore, PHLMM emerges as a "convenient and universal" UHC framework for public health, which is in accordance with the "Sustainable Development Goals 2030" of WHO \cite{who2022universal} and China \cite{China2023}.

% 健康中国行动——心脑血管疾病防止行动实施方案（2023-2030）
Furthermore, the CBPH can function as a Primary Health Care (PHC) system, enabling effective monitoring and early warning of the public health issues \cite{hanson2022lancet, kasper2023rethinking}. As illustrated in Fig. \ref{Fig:PHLMM_overview}, CBPH is a convenient and universal system designed to harness the capabilities of surrounding devices for measuring the physiological signals of the entire population. On one hand, with the escalating global burden of non-communicable diseases, CBPH can serve as a vital integral component of the public healthcare system for both preventive measures and the seamless coordination of lifelong management of chronic diseases. This technology provides substantial benefits in detecting AF compared to usual care and has the potential for broad applicability due to its wide availability of ordinary smartphones \cite{rizas2022smartphone}.
On the other hand, it can also function as a tool for monitoring large-scale infectious diseases \cite{frieden2023road, lal2023primary}, enabling patients' self-management, risk prediction and early warning, and providing decisive support for preventing development trends. 
Therefore, CBPH is poised to play a crucial role in advancing Universal Health Coverage (UHC) \cite{world2023assessing, pablos2023universal} and preparing for "Disease X" in public health monitoring and warning \cite{ramgolam2023preparing, weforum2024DiseaseX}. As AI software, it relies exclusively on existing smart devices and network infrastructure, thereby ensuring greater accessibility and opportunities for individuals across various geographical locations and socioeconomic statuses. CBPH emerges as a "convenient and universal" UHC framework for public health, which is in accordance with the "Sustainable Development Goals 2030" of World Health Organization (WHO) \cite{who2022universal} and China \cite{China2023}.

\section{Database Layer}

This section introduces the VBPM technology to extract physiological information and the LSPH database.

% In this section, we outline the function of each sub-module of CBPH, and the significance of medical scientific discoveries. As shown in Fig. \ref{Fig:PHLMM_overview}, CBPH consists of VBPM technologies, the LSPH database, and the PHAI model. 
% Firstly, VBPM is capable of extracting basic physiological information from videos uploaded by users. Then, LSPH database includes public baseline data, pathology information related to illnesses and clinicians' diagnosis outcomes. Finally, PHAI is a comprehensive multi-task large medical model designed to fulfill distinct roles for patients, AI-developers, and decision-makers by utilizing deep neural networks. 

\subsection{Visual-based physiological monitoring technology}

\begin{figure}[h]
	\centering
	\includegraphics[width = 16.0cm]{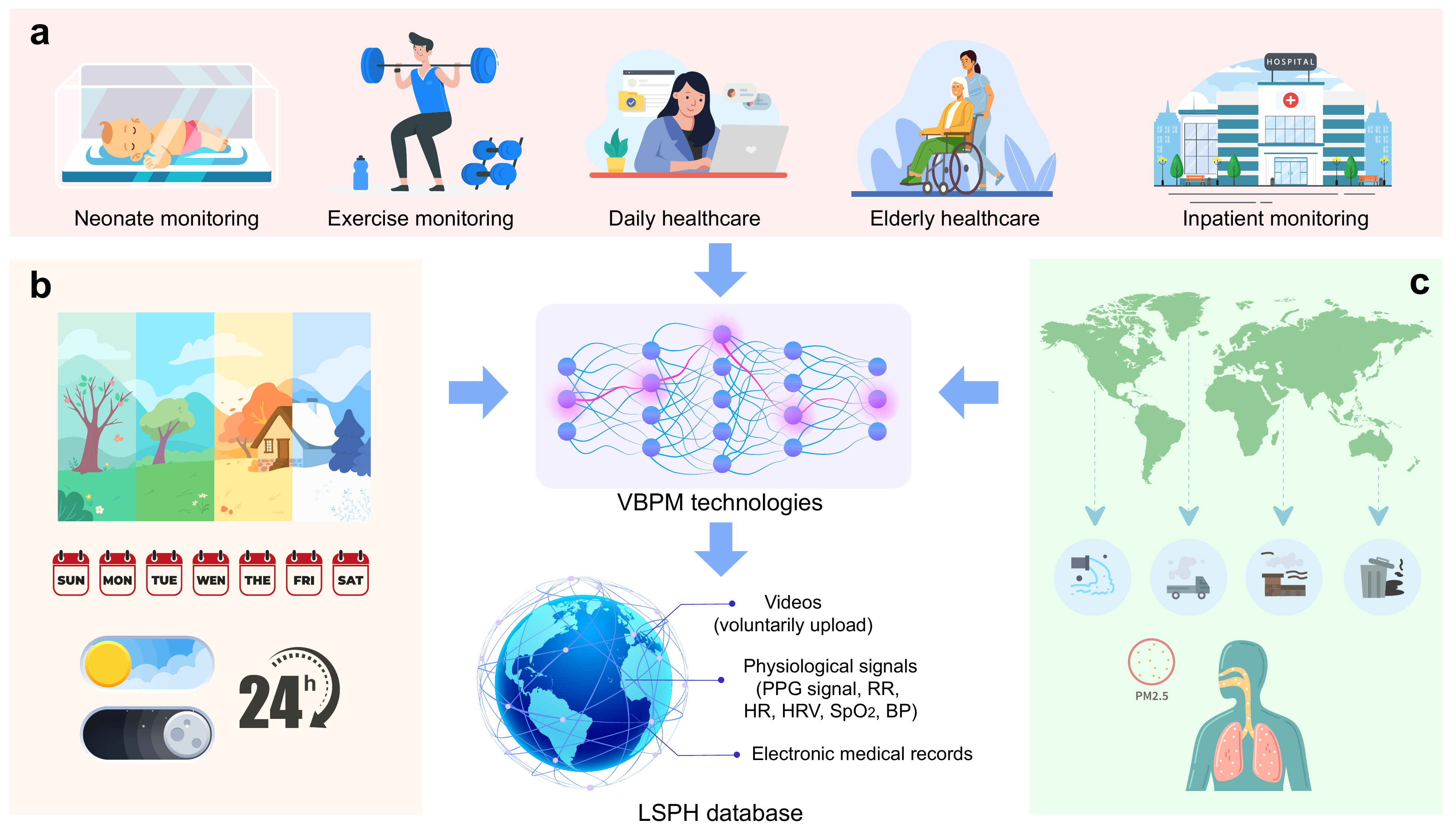}
	\caption{An overview of VBPM technologies for monitoring multiple physiological parameters. VBPM can be conveniently employed in various  application scenarios, times, and regions. \textbf{a}, It can be conveniently applied to diverse populations and scenarios. \textbf{b}, 
		It has the ability to collect physiological information data at any time, and these physiological signals are collected across weeks and seasons. \textbf{c}, The potential for data collection across regions and countries can provide a data foundation for studies on the correlation between regional environments and related diseases.
	}
	\label{Fig:Data_VCPM}
\end{figure}

VBPM is a versatile and flexible solution for healthcare and physiological signal monitoring, including both contact-free methods and smartphone-based fingertip monitoring.
As shown in Fig. \ref{Fig:Data_VCPM}, the VBPM is a personal health technology that has easier access to vital-sign monitoring anywhere and anytime compared with conventional medical devices. It solely utilizes an off-the-shelf camera, such as webcams or smartphones' built-in cameras. On one hand, it can be utilized to establish the physiological benchmarks of individual health along with daily, weekly and seasonal fluctuations \cite{radin2021hopes}. On the other hand, due to the convenient measurement manner, VBPM is versatile and capable of applying to various demographics and scenarios, such as neonates and the elderly. Consequently, VBPM will become the cornerstone for constructing a complex, large-scale physiological information dataset that encompasses the entire population, entire lifecycle, and multiple scenarios.

%The VBPM is one of the crucial components in the PHLMM system, playing an essential role in furnishing users and the PHAI model with physiological parameters. 
In order to safeguard privacy, the VBPM can be implemented on terminal devices such as smartphones, webcams and laptops. Additionally, users can choose whether to upload the raw facial videos to cloud storage based on their specific circumstances. As illustrated in Fig. \ref{Fig:Data_VCPM}, VBPM can be utilized to establish a LSPH database, which spans distinct seasons (timing), regions/countries (location) and demographics (human resources). Thus, LSPH database integrates information from multiple dimensions, including longitudinal individuals' long-term vital signs baseline, and more horizontal relationships with various crowds. Moreover, these characters of VBPM align perfectly with the plan put forward by The Lancet Global Health Commission in 2022 to "advance a human-centered approach of PHC" \cite{hanson2022lancet}.

\subsection{Large-scale public health database}

As shown in Fig. \ref{Fig:PHLMM_overview}\textbf{c}, the LSPH database encompasses three distinct typological and modal datasets: 1) User-uploaded raw data, e.g., videos, on a voluntary basis; 2) Physiological information extracted through VBPM technology; 3) Pathological data contributed by clinicians or obtained from electronic medical records. To ensure the quality of raw data and to improve the reliability of physiological information extracted by VBPM, CBPH not only provides professional raw data recording tools but also establishes standardized operating procedures for collection of raw data. Furthermore, the backend integrates a data quality assessment algorithm to quantitatively ensure the recording quality of raw data. % Undoubtedly, as a privacy safeguard, the raw data recording process occurs locally on the mobile device. Whether to upload the final raw data to the PHLMM cloud storage server is entirely at the discretion of users. 

The vast majority of the LSPH database information is constructed by VBPM algorithm deployed on user terminal computing equipment. As depicted in Fig. \ref{Fig:Data_VCPM}, in the course of users' interaction with the CBPH system, regardless of time or location, the system will systematically extract and archive vital-sign information associated with users' distinct physiological status, such as waking up in the morning, pre and post-exercise periods, and different stages of illness. Compared to datasets constructed based on additional medical devices, the LSPH database not only comprises data covering a broader range of populations and regions, particularly including  LMICs, but also contains tens of thousands of times more data than traditional similar databases. This characteristic enables the LSPH database to form an unbiased, large-scale medical benchmark. Consequently, the LSPH can emerge as a large-scale public health database containing equitable physiological data and developing fair AI models. Moreover, it can be promptly updated by the public, making it a valuable resource for the response to extensive public health events and advancements in medical science.

% In LSPH database, the pathological information serves as the definitive ground truth (label) for predicting and diagnosing illness outcomes. For example, continuous monitoring of a user's cardiorespiratory status enables the AI model to predict the likelihood of the user developing a specific cardiorespiratory disease based on a substantial amount of physiological signal information collected during earlier stages. As a result of the widespread adoption of VBPM technology, LSPH will emerge as a large-scale public health database containing equitable physiological data and developing fair AI models. Moreover, it can be promptly updated by the public, making it a valuable resource for the response to extensive public health events and advancements in medical science.

% Consequently, due to popular of CPM technology, LSPH is a great-scale public health database with fairness physiological data, and can be updated by public in timely. It serve as a database for massive public health events and medical scientific discoveries.

%\textbf{Build fairness database and mitigate algorithm bias.} \cite{chen2023algorithmic}
%
%Firstly, due to the fact that VCPM only requires an off-the-shelf camera and enables contactless measurement, it boasts cost-effectiveness and user-friendly passive monitoring. 
%This capability allows for the convenient, continuous, and real-time monitoring of multiple vital signs, thereby providing more accurate and comprehensive data.

\section{AIaMD Layer}

Data-driven artificial intelligence (AI) models in healthcare and medicine have made exciting advancements \cite{rajpurkar2022ai, yang2022artificial, haug2023artificial, brownstein2023advances, di2023explainable, klauschen2024toward, chen2023digital}. Therefore, governments have actively developed policies or guidelines for AI in the medical field. For instance, both the United States and the European Union have released Acts for AI as a Medical Device (AIaMD) \cite{USFDA2021, madiega2021artificial}.
AIaMD has permeated public health, and played a pivotal role in such areas as disease diagnosis, Internet of Medical Things (IoMT), risk stratification and resource allocation \cite{haug2023artificial, arora2023value}. However, AI medical algorithmic bias and healthcare inequality \cite{ferryman2023considering, irizar2024disproportionate, jones2024causal, garin2023medical, ganapathi2022tackling, mittermaier2023bias, nazer2023bias, seyyed2021underdiagnosis, arora2023value, lancet2023ai} remain the foremost concern for policymakers, medical researchers, AI developers, physicians, and patients. The CBPH framework is devoted to addressing these challenges by significantly increasing the number of users and improving coverage in healthcare, disease diagnosis, and more.

% \subsection{Internet of medical things}
% Clinical applications and IoMT

\subsection{Disease diagnosis}

The COVID-19 pandemic has accelerated the advancement of digital health technology (DHT), leading to exponential growth in camera-based disease diagnosis and detection, particularly using smartphone cameras. These applications include AF screening, hypertension and heart failure monitoring, SpO$_2$ measurement \cite{frey2022blood} and more.
An expanding body of clinical evidence indicates that smartphone camera-based PPG signal for AF detection demonstrates high reliability \cite{rizas2022smartphone, gruwez2023smartphone, khan2023remote, gruwez2024real, fernstad2024validation}. For instance, the study of Rizas et.al demonstrates that smartphone-based PPG signals can be effectively utilized to detect AF and has more than doubled the detection rate of treatment-relevant AF in both phases of their trial \cite{rizas2022smartphone}. Thanks to the convenience and repeatability of camera-based PPG signal measurements, Henri et.al indicates that PPG-based detection of atrial fibrillation (AF) demonstrated outstanding performance with a sensitivity of 98.3\%, specificity of 99.9\%, positive predictive value of 99.6\%, and negative predictive value of 99.6\% \cite{gruwez2024real}.

\begin{figure}
	\centering
	\includegraphics[width = 15.0cm]{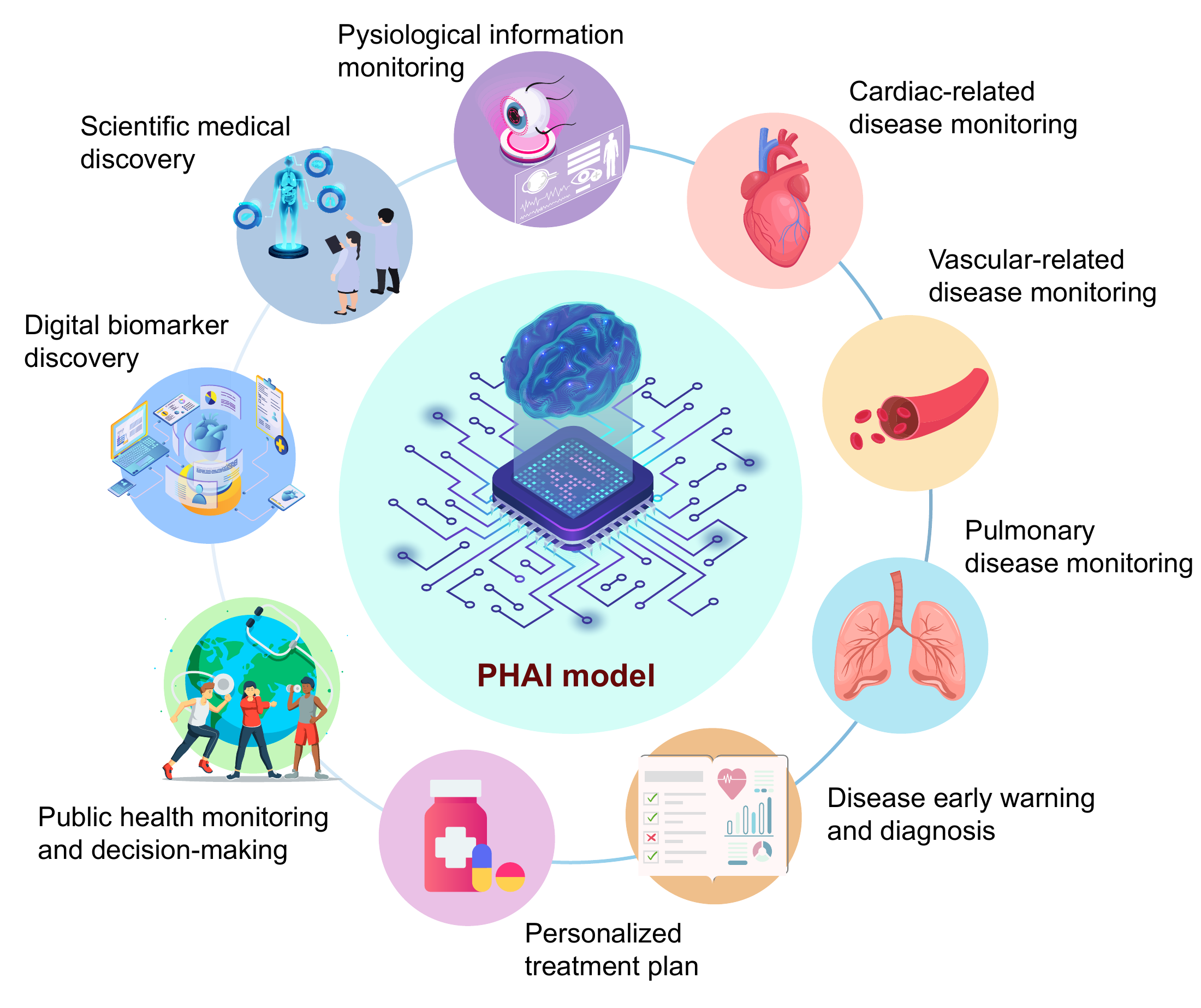}
	\caption{Illustration of the PHAI model showcases its versatility and exciting potential applications across various domains. PHAI can be employed to monitor the progression of public health emergencies, facilitate the healthcare of aging population, forecast and diagnose illnesses like cardiovascular disease, conduct large-scale dataset-driven medical scientific discoveries, and more.}
	\label{Fig:PHAI}
\end{figure}

The utilization of big data and AI technology in medicine and healthcare is an inevitable trend in future development \cite{haug2023artificial, chen2023digital, lancet2023ai, rajpurkar2022ai}. 
As illustrated in Fig. \ref{Fig:PHLMM_overview} and \ref{Fig:PHAI}, the PHAI serves as the core AI fundamental model within the CBPH framework. It can be applied to a wide range of downstream tasks, including cardiopulmonary disease surveillance, personalized treatment planning, scientific medical discoveries, and so on. The PHAI model is a comprehensive multi-task large medical model designed to fulfill distinct roles for users/patients, AI-developers, medical researchers and decision-makers by utilizing AI technologies. The detailed potential applications/functions of CBPH are elaborated in Subsection \ref{sect:healthcare} and Section \ref{sect:AI4Med}. In the realm of individual health, PHAI possesses the capability to model and extract personalized features related to pathology that are concealed within intermittent physiological information across time. On a broader scale, for the public, PHAI has the ability to monitor various cardiorespiratory diseases, identify common characteristics (biomarkers), trace the evolution of these diseases, provide diagnostic support, and formulate personalized therapeutic regimens. Most importantly, once the CBPH system becomes widespread, it will hold strategic importance in monitoring the development trends of sudden public health events and providing decision-makers with quantified data support.

% \section{Discussion}
%  \subsection{Perspective of individuals and public}

% \subsection{Longitudinal and horizontal healthcare}
\subsection{Healthcare}
\label{sect:healthcare}

Longitudinal historical physiological information holds significant research value for personal healthcare and early disease warning \cite{radin2021hopes}. VBPM technology can provide richer data, enabling AI models to gain insights beyond those available from clinic visits.
Given the unique physical conditions of each individual, the collection and research of historical physiological data are indispensable for advancing effective healthcare \cite{radin2021hopes}. On one hand, since VBPM is a versatile and convenient physiological monitoring manner, the CBPH is applicable across various life stages, encompassing surveillance for premature infants/newborns, and health monitoring and assessment for both adults and the elderly. On the other hand, CBPH can also play a pivotal role in monitoring the variations in physiological information throughout the entire course of a disease, such as early disease detection, interventional treatment (mid-term), and late-stage rehabilitation.

\textbf{Personal monitoring technology for PHC.}
PHAI can serve as a personal monitoring technology forming the cornerstone of a high-quality healthcare system, such as the screening and early detection of diseases \cite{croke2024primary, brownstein2023advances}. For each patient/individual, they are not only the participants that build the LSPH database, but also capable of obtaining themselves the feedback of physiological information from VBPM.
The CBPH utilizes personal monitoring technologies to achieve PHC system. It leverages the personal historical baseline (individualized characteristics), current pathology information, and horizontal relationships with other patients related to the disease (common characteristics) to optimize a therapeutic regimen tailored to the individual. Overall, this system facilitates a deeper understanding of one's health status, timely detection of health issues, and proactive measures or medical intervention. The integration of personal monitoring technologies with PHC enhances the efficiency and quality of medical services, reduces the incidence and progression of diseases, and enhances overall health outcomes \cite{hanson2022lancet, brownstein2023advances}. 

\textbf{Public health and decision-making system}.
The number of built-in smartphone cameras will facilitate the deployment of CBPH system infrastructure on a scale surpassing any other type of healthcare equipment, including wearable devices. This widespread adoption of CBPH systems is expected to bridge socioeconomic and regional disparities, significantly advancing global sustainable development and health well-being. Thus, CBPH can be employed to predict public health outcomes based on extensive physiological information during responses to public health emergencies, including the identification of outbreaks of infectious diseases that may impact public health \cite{haug2023artificial}. Furthermore, CBPH can be considered as a decision-making system for analyzing and predicting the affected population in response to public health emergency. Its primary objective is to furnish decision-makers with more precise information of public health events, thereby enhancing strategic planning and subsequent response efforts, which is particularly crucial in addressing highly transmissible or lethal diseases, such as "Disease X" \cite{ramgolam2023preparing, weforum2024DiseaseX}. Consequently, through the CBPH system, decision-makers can promptly access information on disease transmission, health risks, and allocation of medical resources, enabling them to formulate more effective public health policies and action plans in advance. As a whole, CBPH can improve public health, reduce the burden of disease, and promote the health and well-being of society.

\section{AI4Medicine Layer}
\label{sect:AI4Med}

Due to innovations in sensor technology, including advancements in remote medicine and the Internet of Medical Things (IoMT), data-driven medical scientific discoveries have made significant progress in recent years. However, the population covered by uniform data remains limited, and its scale is insufficient for training large language models effectively.

\subsection{Precision medicine}
% precision medicine and disease evolution process

Physiological signs are manifestations of physical conditions and are crucial for diagnosing diseases, monitoring health status, and evaluating treatment effectiveness \cite{spatz2024wearable,bramante2023outpatient, clarke2023factors}. Fluctuations in these essential indicators often occur during the latent and acute phases of numerous diseases, as well as in response to abnormal physiological conditions, potentially indicating the underlying pathological process \cite{radin2021hopes}. Clinicians commonly measure these vital signs to assess the current physiological status and discern trends in the progression of a patient's condition. However, this approach is inherently empirical and cannot quantify or precisely assess the evolution of diseases within the human body. Therefore, long-term quantification of physiological information is of great values in AI4Medicine for monitoring disease progression, evaluating drug responses, and optimizing treatment plans, especially in dealing with unknown diseases. Through AI modeling of large medical dataset and considering the distinction in each individual's physical condition (e.g., having different underlying diseases) and drug resistance response, CBPH can discover new disease pathogenesis, diagnostic methods, and personalized treatment plans, thereby advancing AI4Medicine.

\begin{figure}
	\centering
	\includegraphics[width = 16.0cm]{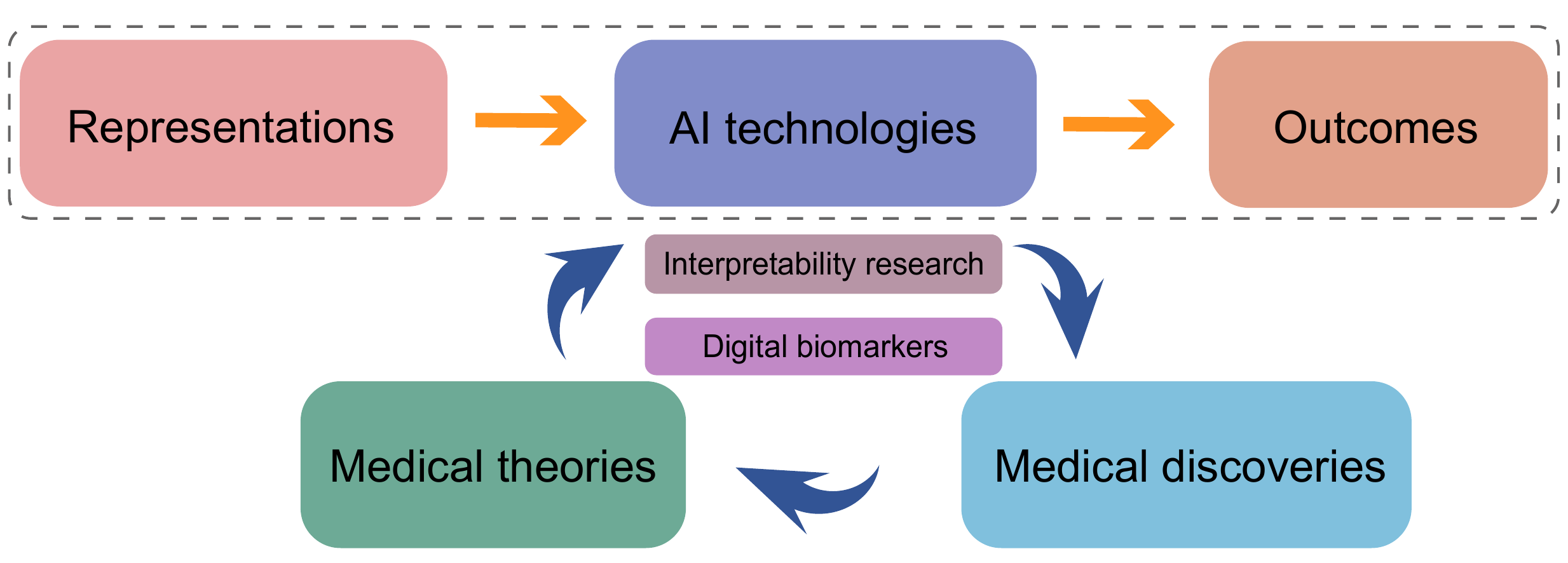}
	\caption{Data-driven AI technologies for medical scientific discoveries.  Traditional AI medical technologies model from representations to outcomes. The CBPH system emphasizes its role in facilitating the transition from new medical scientific discoveries based on large-scale data-driven AI technology to novel medical theories. It establishes a positive cyclical research process between "data-AI technology-medical theory" based on the studies of interpretability and discoveries of digital biomarkers.}
	\label{Fig4:MedScience}
\end{figure}

The development of AI4Medicine and public health is an inevitable trend \cite{haug2023artificial, dicuonzo2023healthcare, lancet2023ai, rajpurkar2022ai}. However, how to achieve this, how to regulate it, and how to improve the quality and efficiency of medical care are still urgent problems that need to be addressed. Apparently, it is imperative to construct a more robust and interpretable LMM by leveraging the professional medical knowledge provided by clinicians. As illustrated in Fig. \ref{Fig4:MedScience}, CBPH can not only accomplish traditional AI medical modeling from representations to outcomes, but also facilitate the transition from new medical scientific discoveries based on AI technology to novel medical theories. These theories should then be used to further enhance the performance of PHAI model. Thus, this approach will establish a positive cyclical research process between "data-AI technology-medical theory".
Furthermore, CBPH not only enhances the model's generalization capability and interpretability, but also facilitates the exploration and development of innovative and undiscovered medical theories. For instance, it can discover the relationship between new digital biomarkers \cite{lan2023performer, avram2020digital, saner2020contact} derived from physiological information and the symptom development of one specific disease. The biomarkers have the potential to revolutionize healthcare by providing new ways to monitor and diagnose diseases. Thus, by AI modeling of large-scale physiological data, these discoveries can lead to more accurate disease diagnosis and prognosis, optimize personalized treatment plans, and improve healthcare management strategies.

\subsection{Medical scientific discovery}

Based on the large-scale public physiological information from the aforementioned horizontal and longitudinal data, PHAI model can be further developed to track a disease evolution process.
This tracking holds significant value, especially in the research of intractable or unknown diseases, such as cancer\cite{bailey2021tracking}, cardiovascular disease \cite{zwack2023evolution, spatz2024wearable} and COVID-19 (or Disease X) \cite{brownstein2023advances, fawzy2023clinical, sudat2023racial, bramante2023outpatient, clarke2023factors}. As highlighted in the study of Spatz et al. \cite{spatz2024wearable}, a 62-year-old woman with long-standing hypertension and AF requires vital-sign monitoring from early warning to post-hospital rehabilitation to develop personalized therapy procedure. Generally, for most diseases are closely related to fundamental vital signs, long-term physiological signal monitoring is of significant importance for tracking the evolution of diseases and formulating treatment plans. Therefore, CBPH system can manifest considerable clinical research potential in the establishment of large-scale physiological information databases, disease prediction and diagnosis, as well as personalized treatment plan development (e.g. prognosis and treatment response) \cite{spatz2024wearable, bailey2021tracking}.

% \subsection{Perspective of AI-developers and clinicians}

% Heart rate, respiration, body temperature, blood oxygen, and blood pressure are recognized as the five fundamental vital signs in humans. 
Recently, there has been an abundance of novel medical scientific discoveries based on AI technology. Typical recent discoveries related to physiological information include but are not limited to smartphone camera-based AF screening \cite{rizas2022smartphone, ding2023log, gruwez2023smartphone}, COVID-19 patient condition diagnosis based on HRV \cite{mol2021heart, hasty2021heart}, diabetes monitoring based on smartphone PPG signals \cite{lan2023hdformer, avram2020digital}, and more \cite{lan2023performer, charlton2022assessing, weng2023predicting, JAVAID2022100379}. These new discoveries demonstrate the broad application prospects of the CBPH system in the medical field, providing new possibilities for improving health monitoring and disease management.

% \cite{croke2024primary}

%\subsection{Progress toward the SDGs}
%% \textbf{Camera sensor VS wearable device.}
%
%Good Health and Well-being (SDG 3) is the third goal of the SDGs \cite{UN2020SGD3}. In September 2023, the WHO and the World Bank released the "Tracking UHC: 2023 global monitoring report," which stated that in 2021, approximately 4.5 billion people (more than half of the global population) did not have adequate access to essential health services \cite{world2023tracking}. Additionally, Dr. Tedros, Director-General of the WHO, stated, "The COVID-19 pandemic reminds us of the critical importance of providing primary health care services. Healthy people are the foundation of a healthy society and economy." PHLMM is crucial for health monitoring in LMICs as it can reach more people while maintaining cost-effectiveness. This is essential for promoting the SDGs.

% \textbf{Personal monitoring technology for PHC}.
% 基于个人监测技术的初级医疗保健系统是指利用先进的监测技术，如智能手环、智能手机应用程序等，对个体的健康状况进行实时监测和评估，并基于监测结果提供个性化的医疗保健建议和服务的系统。这种系统可以帮助人们更好地了解自己的健康状况，及时发现健康问题，并采取预防措施或寻求医疗帮助。通过将个人监测技术与初级医疗保健相结合，可以提高医疗服务的效率和质量，减少疾病的发生和发展，提高人们的健康水平。

% 构建无偏大数据集，训练医疗大模型，探索基于大模型的医学理论与可解释性研究

\subsection{Evidence-based medicine}

% digital biomarker and endpoint for EBM

%The CBPH framework can promote the development of data driven evidence-based medicine, including digital biomarker discovery, digital endpoint 
%The VBPM technology can enable digital endpoints in any
%Offering large-scale data for studies and clinical trials to discover digital biomarkers. VBPM can contribute to digital endpoints by providing continuous, non-invasive, and real-time physiological data, thus enhancing the precision of these assessments. Thus, CBPH framework holds great promise for advancing evidence-based medicine by providing precise, continuous, and non-invasive monitoring of vital signs.

The VBPM technology and LSPH database can provide the foundational  physiological data for advancing digital biomarkers, digital endpoints and evidence-based medicine (EBM). The VBPM is an essential complement to traditional medical data sources, providing additional insights into the patient's daily physical condition that may not be captured during clinician visits. It can discover new digital biomarkers and endpoints that correlate with disease progression and treatment response. As Subbiah's perspective paper point out that sensors and IoMT technology offer many opportunities to acquire data and advance the next generation of EBM \cite{subbiah2023next}. Digital endpoints are clinical outcome assessments (COAs) derived from digital biomarkers, often used in clinical trials to measure the effect of a therapeutic intervention.  
Its ability to generate valuable digital biomarkers and endpoints benefits all stakeholders by leading to better patient outcomes and more efficient clinical trials. This, in turn, accelerates drug development, promotes cost-effective personalized healthcare, and improves public health.

%Its ability to generate valuable digital biomarkers and endpoints for all stakeholders, which can lead to better patient outcomes, more efficient clinical trials, accelerating drug development and promoting cost-effective, a more personalized approach to healthcare, and improving public health.

\subsection{Large medical model}

\textbf{Large Physiological Model (LPM)}
The success of LLMs is attributed to their effective utilization of massive datasets, and similar strategies can be applied in the public health domain. Consequently, the CBPH framework, capable of collecting physiological data from billions of users worldwide, is highly likely to construct a LPM for public health in the future. First, due to the prevalence of smartphones and other camera devices, the CBPH framework is highly feasible and widespread. This facilitates large-scale data collection and analysis, encompassing users globally, and providing a solid foundation for building extensive datasets. Second, physiological information exhibits significant temporal characteristics, similar to language/text data. By constructing large models capable of capturing the patterns of physiological changes, it becomes possible to achieve early disease warning, diagnosis, and personalized treatment. Third, the collection and analysis of large-scale physiological data will propel the advancement of medical research, particularly data-driven medical research. Through in-depth mining of these data, new biomarkers can be discovered, and the mechanisms of disease occurrence and development can be unveiled, thus offering new insights and methods for disease diagnosis and treatment. Therefore, the CBPH framework holds substantial potential and significance in constructing medical large models for public health. In the future, with the development of technology and the increase in data, the LPM is expected to play an increasingly vital role in the public health domain.

\textbf{Multimodal Medicine Large Language Model (MM-LLM)}
In addition, existing large medical language models are predominantly question-and-answer systems based on medical texts or dialogues. While textual information carries a certain degree of subjectivity, the physiological signals recorded by CBPH system are more objective. Additionally, physiological data provide critical historical information of personal health. It is indisputable that a MM-LLM integrated physiological information with textual data can significantly enhance its performance. The optimization of the MM-LLM by incorporating extensive physiological data can greatly improve their capabilities and application range. This integration not only enhances the accuracy of the model in answering medical-related questions but also makes it more practical for diagnostics, prediction, and personalized medical advice. By leveraging both objective physiological signals and the contextual richness of textual data, the advanced model can offer a more comprehensive and accurate representation of users' physical condition. For example, in June 2024, Cosentino et al. (Google) and Fang et al. (MIT) were the first to integrate physiological information from wearable devices into large language models, pioneering research on large models for personal health \cite{cosentino2024towards, fang2024physiollm}. 
Consequently, the development and implementation of the MM-LLM that synergizes physiological and language information can revolutionize medical diagnostics and personalized healthcare, providing more reliable and precise tools for clinicians and researchers. This approach represents a significant step forward in the creation of robust, data-driven medical models that bridge the gap between subjective and objective health information.

\section{Usage cases}

In this section, two usage cases of the CBPH system are depicted, namely, responding to fulfilling healthcare requirements of the aging population and the COVID-19 pandemic.

\subsection{Healthcare for non-communicable diseases of aging population}

Non-communicable diseases (NCDs) claim the lives of 41 million people annually, constituting 74\% of global deaths \cite{WHO2023ncds}. Among these NCD-related fatalities, 77\% of these mostly premature and avoidable deaths occur in low- and middle-income countries \cite{WHO2023ncds, mohan2023principles}. Furthermore, cardiovascular diseases are responsible for the majority of NCD deaths, affecting about 17.9 million people annually \cite{WHO2023ncds}. Additionally, the global elderly demographic is undergoing substantial expansion, with individuals aged 65 years or above expected to increase from 10\% (771 million) in 2022 to 16\% (1.6 billion) in 2050 \cite{WorldPopulation2022}. Consequently, the WHO has introduced two key initiatives: a "Global Action Plan" to cut NCDs premature mortality by 25\% by 2025, and SGDs for 2030 with a target to reduce premature NCD-related deaths by 30\% \cite{bennett2020ncd}.

In response to the rapid acceleration of global aging and the monitoring and prevention of cardiovascular diseases, the CBPH system, as it does not require additional specialized medical hardware support, remains unchallenged with the increasing number of patients. The long-term monitoring of vital signs in the elderly can serve as an early warning system for emergency events and deteriorating health conditions, including the early onset of cardiovascular \cite{rizas2022smartphone}, neurological, and pulmonary diseases \cite{chen2023digital}. Most importantly, it holds the potential to eliminate healthcare disparities and structural inequalities in medical access for elderly individuals in LMICs. The commentary article of Nature Medicine in 2023 highlights the "Principles for NCD prevention and control," with two of the five principles being: 1) Monitoring and evaluating risks throughout the life course. 2) Providing human-centered care based on digital health technologies and non-physician healthcare personnel \cite{mohan2023principles}. These principles align closely with the original design of CBPH, further confirming its immense potential for applications to monitoring and preventing cardiovascular diseases in the elderly. 
In summary, CBPH's capacity for convenient monitoring of cardiovascular disease-related physiological parameters in the elderly, anytime and anywhere, is pivotal for early detection, warning, and long-term post-discharge monitoring of cardiovascular diseases \cite{spatz2024wearable}. This is expected to significantly contribute to preventing more cases of premature mortality.

% 早，中，晚，每个时期的作用

%\begin{figure}
%	\centering
%	\includegraphics[width = 16.0cm]{Fig5_COVID_19.pdf}
%	\caption{The functions and roles of CBPH in response to COVID-19.}
%	\label{Fig:covid19_case}
%\end{figure}

\subsection{Preparing for Disease X or the COVID-19 Pandemic}

% From 2019 to 2023, people were in panic due to the rapid spread of COVID-19, which was a protracted and severe war for all people. Take the epidemic in Japan, USA and France as an example. About 10 waves of the epidemic swept these countries successively within three years. Thanks to strict control measures, the world's case fatality rate decreased from 7.70\% in early 2022 to 1.08\% in September, 2022. However, compared to the common flu, the case fatality rate was still ten times higher (https://ourworldindata.org/). Furthermore, even in recent days, according to incomplete statistics, over 850,000 new COVID-19 cases were reported during the 28-day period from 20 November to 17 December, 2023, which is a 52\% increase than that in the previous period \cite{world2023covid}.
% From 2019 to 2023, people were in panic due to the rapid spread of COVID-19, which was a protracted and severe war for all people. Compared to the common flu, the case fatality rate was still ten times higher (https://ourworldindata.org/). Furthermore, even in recent days, according to incomplete statistics, over 850,000 new COVID-19 cases were reported during the 28-day period from 20 November to 17 December, 2023, which is a 52\% increase than that in the previous period \cite{world2023covid}.

Since the global outbreak of the COVID-19 pandemic in 2019, SARS-CoV-2 has persisted, spreading seasonally rather than disappearing. For example, in the summer of 2024, countries such as Japan and the United States experienced new waves of COVID-19 outbreaks. Moreover, SARS-CoV-2 has been continuously mutating, with virologists diligently monitoring and analyzing the transmissibility and lethality of emerging variants.
Therefore, experts have advocated for the universal application of personal health technologies \cite{radin2021hopes} and remote monitoring systems. Greenhalgh et al. found that pulse oximeters could detect hypoxia associated with acute COVID-19 at home \cite{greenhalgh2021remote}, which indicated the possibility of home monitoring during the pandemic. Moreover, in response to the COVID-19 outbreak, the WHO released the interim guidance for member states on the use of pulse oximetry in monitoring COVID-19 patients under home-based isolation and care \cite{world2021interim}, in order to decrease unnecessary exposure of seeking medical advice. It can be seen that the 24-hour monitoring of physiological parameters plays an essential role in examining the coronavirus. New mobile and telemedicine technologies might help patients and their families to take greater responsibility for their own health and enable new and more horizontal relationships between patients and providers \cite{hanson2022lancet}.

\begin{center}
	\begin{tcolorbox}[colback=green!10, %gray background
		colframe=black,% black frame colour
		width=16cm,% Use 8cm total width,
		arc=1mm, auto outer arc,
		boxrule=0.5pt,]
		\centering{\large{\textbf{Box1: Response to COVID-19 pandemic equally applicable to Disease X} \\}}
		~\\
		\hrule
		~\\
		
		% \large{ \textbf{Perspective of medical researcher and decision maker \\}}

		\begin{itemize}[leftmargin=0.3cm] %itemindent=1cm]
			\item Early stage: The rapid infection capability of SARS-CoV-2 and insufficient understanding. CBPH can rapidly gather a substantial amount of vital sign information from patients at each stage of post-COVID-19 infection. This enables the exploration of infection and evolution of SARS-CoV-2 within the body based on physiological signals, providing the support of extensive quantitative data for determining appropriate treatment plans and advancing scientific research and medical understanding of the virus. Therefore, CBPH can facilitate scientific discoveries related to medical knowledge based on data science. 
			
			\item Middle stage: Through the early development and refinement of the CBPH model, it can be further trained to serve as a diagnostic model for COVID-19 infections, with triage capabilities. It can discern patients who merely need home isolation treatment, those requiring ventilator support, and those necessitating the transfer to intensive care units. Consequently, in the scenario of a widespread infection, CBPH can proficiently allocate valuable medical resources, optimize medical efficiency, and thus save more lives.
			
			\item The post-Covid era: Although the COVID-19 has been reclassified as a "Class B communicable disease monitored as category B", we still cannot take it lightly since there remain several highly infectious variants such as the JN.1. The conventional medical system faced great pressures at the advent of the pandemic due to the surge in the number of patients, the inevitable infection of medical personnel and the shortages of medical resources. CBPH is a primary health care and universal health coverage framework for public health, enabling real-time monitoring and early warning of the next epidemic. % Certainly, The PHLMM can also serve as a framework for responding to other public health events.
			
		\end{itemize}
		
	\end{tcolorbox}
\end{center}

However, individuals experience daily, weekly, and seasonal fluctuations unique to them in a range of physiological parameters and activities. Only by knowing the long-term and normal physiological state of an individual can the earliest deviations be identified precisely. As such, CBPH can be regarded as a remarkable way in the control of sudden public health events such as COVID-19 \cite{brownstein2023advances}. Specifically speaking, with VBPM technology, people can record their own facial videos and get the physiological parameter feedback whether in the pandemic or not. Simultaneously, people can upload their physiological data to cloud willingly to facilitate the public health prevention. Then, with such large-scale and diverse physiological data from people of different ages and regions, the unbiased LSPH database can be constructed successfully. Next, the characteristics of the coronavirus and the symptoms of the disease can be analyzed by adopting the mature PHAI model with the LSPH database. Thus, it can be accurately and timely to distinguish between patients who can be managed safely in the home settings and those who must be sent to hospital \cite{espinosa2022remote}, which can greatly alleviate the pressure of hospitals and optimize the allocation of medical resources. In addition, people can know the course of disease quantitatively from the results of big data rather than speculate their own conditions by contrasting symptoms only from the Internet shared from the public.

It is important to note that although the peak of the pandemic has passed, the novel coronavirus has not been completely eradicated and there still exists potential epidemic risks. Therefore, CBPH still bears significant practical relevance, e.g., monitoring the infection capability and development trends of variant strains of the SARS-CoV-2. In today’s post pandemic era, CBPH can be devoted to summarizing the scientific discoveries, including the coronavirus evolution process in human bodies and the personalized therapeutic schedule for individuals, which can facilitate the development of AI4Medicine. Overall, the advancement of CBPH is needed to align with public health \cite{frieden2023road} and UHC, and we firmly believe that CBPH will show its capabilities in addressing future pandemics \cite{irizar2024disproportionate}.

\section{Discussion and Outlook}
% 从公共健康和AI的角度分别切入，阐述CBPH的优势

AI has indeed become integral to public health, significantly impacting disease diagnosis, risk assessment, and resource distribution \cite{arora2023value}. Nevertheless, the prevalence of bias in AI medical algorithms and the resulting healthcare disparities \cite{ferryman2023considering, irizar2024disproportionate, jones2024causal, garin2023medical, ganapathi2022tackling, mittermaier2023bias, nazer2023bias, seyyed2021underdiagnosis, lancet2023ai} remain paramount concerns for all stakeholders. Additionally, in the context of escalating global aging population and the impact of the post-COVID era on public health, the establishment of CBPH system holds strategic significance and necessity for the advancement of global public health. Firstly, the CBPH boasts a promising prospect as a proper solution to the establishment of a large-scale and cross-country medical database \cite{croke2024primary}, which will overcome the challenges of insufficient data in the PHC system \cite{croke2024primary} and unfair AI medical models \cite{garin2023medical, mittermaier2023bias, nazer2023bias, chen2023algorithmic, arora2023value, lancet2023ai}. This database aligns with the Primary Health Care Performance Initiative Vital Signs Profiles and WHO PHC Measurement Framework and Indicators \cite{vital2023, WHOPHCframework2022}. Secondly, by measuring and comprehensively evaluating public physiological signals through the CBPH system, timely and quantifiable data support can be provided for the research on sudden public health events and policy formulation. For instance, the CBPH framework can be utilized to monitor the development trend of the COVID-19 or "Disease X" \cite{weforum2024DiseaseX} pandemic. Furthermore, based on the LSPH database and LMM, the collaborative study between AI developers and medical researchers will facilitate medical scientific discoveries.

Moreover, the CBPH system will become the first foundational LMM and sustainable development PHC framework, playing a crucial role in addressing infectious public health emergencies and the healthcare requirements of the global aging population, since it will help in the rational allocation of medical resources and the reduction of mortality rates.
Based on VBPM technology, the core idea of CBPH provides a new research paradigm from medical data to medical scientific discovery, offering new perspectives and solutions for the application of AI technology in the field of medicine. It can help AI developers and medical researchers more effectively understand, analyze, and utilize large-scale medical physiological data to discover potential correlations and patterns, thereby promoting the accumulation and innovation of medical knowledge. 
Finally, we firmly believe that CBPH can facilitate a qualitative transformation in the effectiveness and universality of physiological signal monitoring and healthcare. This has the potential to bring about a significant revolution in medical science and AI technology, benefiting public health and AI4Medicine, and promoting the SDGs.

Finally, the CBPH framework will be utilized to address the challenges from the perspective of AI in public health as the following: 
\begin{itemize}
	\item \textbf{Eliminate the bias of medical database and unfairness of AI medical algorithm}. Recently, widespread concerns have emerged in AIaMD due to insufficient data diversity and algorithmic unfairness \cite{ferryman2023considering, garin2023medical, ganapathi2022tackling, mittermaier2023bias, nazer2023bias, chen2023algorithmic, seyyed2021underdiagnosis, arora2023value, lancet2023ai}. The development of AI technology in medicine aims to mitigate medical bias rather than exacerbate it. Hence, the LSPH database and CBPH framework, specifically proposed and designed for public health, are intended to mitigate or potentially eliminate the prevailing algorithmic bias in AI4Medicine and make significant breakthroughs.
	
	\item \textbf{Tackle the challenges posed to public health in the post-COVID-19 era within the situation of globalization}. Due to CBPH's convenient measurement manner for healthcare evaluation and without any requirement for specialized medical equipment, the proposed CBPH framework serves as a PHC and UHC solution for global health to eliminate inequitable healthcare \cite{lal2023primary, hanson2022lancet, frieden2023road, henderson2023understanding, sachs2022lancet}. % In addition, PHLMM is also a effective solution in response to "Disease X" \cite{ramgolam2023preparing}.
	
	% \item The vast potential presented by large-scale medical datasets and models .
	\item \textbf{Leverage the significant potential of large-scale medical database and AI model to accelerate the development of AI4Medicine}. Inspired by the groundbreaking advancements in language models and AI for Science (AI4Science), there is no denying that CBPH plays a crucial role in advancing medical scientific discoveries and deploying AI medical models \cite{singhal2023large, thirunavukarasu2023large, kedia2024chatgpt}. For example, (1) Karan et al. demonstrated in their research on the Large Language Model (LLM) in medical knowledge encoding that as the parameters of the large model increase, its accuracy also increases \cite{singhal2023large}; (2) Avram et al. found that the smartphone-based PPG waveform can provide a readily attainable, non-invasive "digital-biomarker" for prevalent diabetes, based on a study involving 53,870 individuals \cite{avram2020digital}.
	
	\item \textbf{Promote the Sustainable Development Goals (SDGs).} CBPH serves as a PHC and UHC framework, particularly in low- and middle-income countries (LMICs). It has the potential to significantly enhance the effectiveness and efficiency of healthcare in LMICs, contributing to improved health outcomes and promoting the development of the SDGs.
	
\end{itemize}

\section*{Acknowledgments}

This work is supported by the National Natural Science Foundation of China (62203037 and 62271241).
%the National Key R\&D Program of China (2022YFC2407800), Guangdong Basic and Applied Basic Research Foundation (2023A1515012983), and Shenzhen Fundamental Research Program (JCYJ20220530112601003). 

%It is also jointly supported by National Chung Hsing University and Taichung Veterans General Hospital (TCVGH-NCHU 1110104).

\section*{Author contributions}
B.H., C.Z., W.W., and H.L. conceptualized and designed the work. B.H., C.Z., Z.L., S.H., B.Z., H.L., Z.L., W.W. and H.L. discussed the detailed function of CBPH framework. B.H., C.Z., Z.L., S.H., B.Z., H.L., W.W. and H.L. wrote and revised the paper. All authors approved the final manuscript.

\section*{Corresponding author}
Correspondence to Bin Huang (marshuangbin@buaa.edu.cn), Wenjin Wang (wangwj3@sustech.edu.cn), and Hui Liu (liuhui@pumc.edu.cn).

\section*{Data and code availability}
No new data and code were generated in this study.

\section*{COMPETING INTERESTS}
The authors declare no competing interests.

\bibliographystyle{unsrt}
\bibliography{sample-base}

\end{document}